\documentclass[12pt]{elsarticle}
\usepackage{epsfig}
\usepackage{rotating}
\usepackage{lscape}
\usepackage{amsmath}

\biboptions{sort&compress}

\begin{document}

\begin{flushright}

\end{flushright}

\vskip 0.5cm


\begin{center}

{\Large {\bf New limits on $2\varepsilon$, $\varepsilon\beta^+$
and $2\beta^+$ decay of $^{136}$Ce and $^{138}$Ce with deeply
purified cerium sample}}

\vskip 0.5cm

\vskip 0.5cm

{\bf P.~Belli$^{a,b}$, R.~Bernabei$^{a,b,}$\footnote{Corresponding
author. {\it E-mail address:} rita.bernabei@roma2.infn.it
(R.~Bernabei).}, R.S.~Boiko$^{c,d}$, F.~Cappella$^{e}$,
R.~Cerulli$^{f}$, F.A.~Danevich$^{c}$, A.~Incicchitti$^{e,g}$,
B.N.~Kropivyansky$^{c}$, M.~Laubenstein$^{f}$,
V.M.~Mokina$^{c,e}$, O.G.~Polischuk$^{c}$, V.I.~Tretyak$^{c}$}

\vskip 0.3cm

$^{a}${\it INFN sezione Roma ``Tor Vergata'', I-00133 Rome, Italy}

$^{b}${\it Dipartimento di Fisica, Universit$\grave{a}$ di Roma
``Tor Vergata'', I-00133 Rome, Italy}

$^{c}${\it Institute for Nuclear Research, 03028 Kyiv, Ukraine}

$^{d}${\it National University of Life and Environmental Sciences
of Ukraine, 03041 Kyiv, Ukraine}

$^{e}${\it INFN sezione Roma, I-00185 Rome, Italy}

$^{f}${\it INFN, Laboratori Nazionali del Gran Sasso, I-67100
Assergi (AQ), Italy}

$^{g}${\it Dipartimento di Fisica, Universit$\grave{a}$ di Roma
``La Sapienza'', I-00185 Rome, Italy}

\end{center}

\vskip 0.4cm

\noindent {\bf Abstract}

\noindent A search for double electron capture ($2\varepsilon$),
electron capture with positron emission ($\varepsilon\beta^+$),
and double positron emission ($2\beta^+$) in $^{136}$Ce and
$^{138}$Ce was realized with a 465 cm$^3$ ultra-low background HP
Ge $\gamma$ spectrometer over 2299 h at the Gran Sasso underground
laboratory. A 627 g sample of cerium oxide deeply purified by
liquid-liquid extraction method was used as a source of $\gamma$
quanta expected in double $\beta$ decay of the cerium isotopes.
New improved half-life limits were set on different modes and
channels of double $\beta$ decay of $^{136}$Ce and $^{138}$Ce at
the level of $T_{1/2}>10^{17}-10^{18}$ yr.

\vskip 0.4cm


\section{Introduction}

A great interest in double beta ($2\beta$) decay is motivated by
the observation of neutrino oscillations phenomena. The
neutrinoless ($0\nu$) mode of $2\beta$ decay is forbidden in the
Standard Model of particles since the process violates the lepton
number and requires neutrino to be a massive Majorana particle
\cite{Barea:2012,Rodejohann:2012,Delloro:2016,Vergados:2016}. In
addition to the massive neutrino mechanism, there are a lot of
other hypothetical mechanisms of the $0\nu2\beta$ decay
\cite{Deppisch:2012,Bilenky:2015}.

The two neutrino double beta minus ($2\nu2\beta^-$) processes,
allowed in the Standard Model,  are already observed in several
nuclei, while the half-life limits set on the $0\nu$ mode achieved
in the most sensitive experiments are at level of $T_{1/2}>
10^{23}-10^{26}$ yrs (see reviews
\cite{Tretyak:2002,Elliott:2012,Giuliani:2012,Cremonesi:2014,Gomes:2015,Sarazin:2015,
Delloro:2016} and the recent experimental results
\cite{GERDA,EXO-200,CUORE,NEMO-3,KamLAND-Zen}). The experiments
bound the effective Majorana neutrino mass at the level of 0.1 eV
-- a few eV depending on experiment and on the nuclear matrix
elements calculations, which still give a rather wide spread of
values. The experimental sensitivity to the double beta plus
processes: double electron capture ($2\varepsilon$), electron
capture with positron emission ($\varepsilon\beta^+$), and double
positron emission ($2\beta^+$) is substantially lower. Even the
allowed two neutrino mode of these processes has not yet been
observed unambiguously (see reviews
\cite{Tretyak:2002,Maalampi:2013}). At the same time, there is a
strong motivation to enhance the experimental sensitivity to the
$0\nu2\varepsilon$ and $0\nu\varepsilon\beta^+$ decay processes
since these investigations could clarify the possible contribution
of the right-handed currents to the $0\nu2\beta^{-}$ decay rate if
observed \cite{Hirsch:1994}.

For neutrinoless double electron capture, the so-called resonant
mechanism was discussed, when the energy release is very close to
the energy of one of the excited levels of the daughter nucleus
(see \cite{Bernabeu:1983,Kolhinen:2011,Kotila:2014} and references
therein). In the other case, to save the energy-momentum
conservation, it is supposed that the energy excess is taken away
by emission of one or two gamma quanta, conversion electron or
$e^+e^-$ pair (see e.g. \cite{Doi:1993}).

\begin{figure}[htb]
\begin{center}
 \mbox{\epsfig{figure=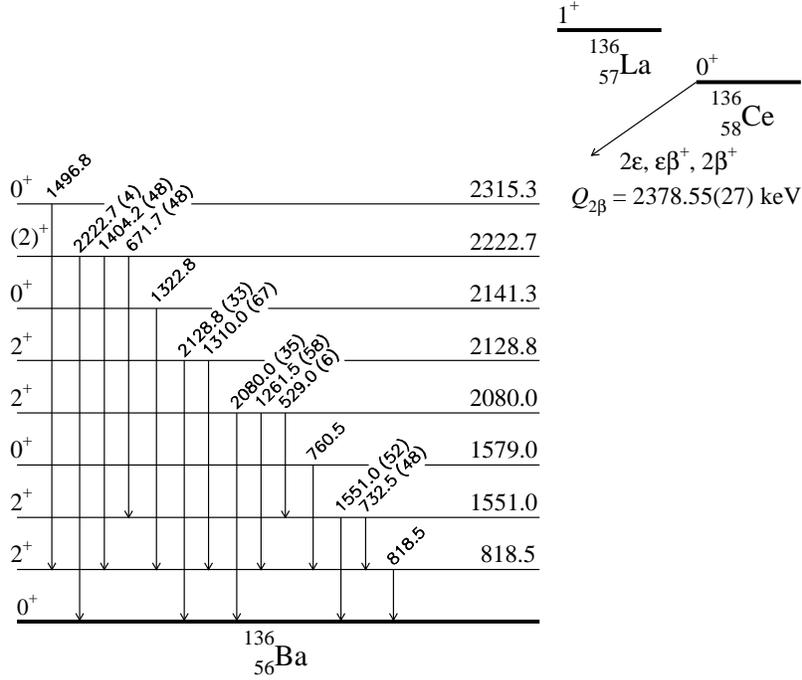,height=9.0cm}}
 \caption{The simplified decay scheme of $^{136}$Ce. Energy of the levels and $\gamma$ quanta are in keV, the
transition probabilities (in \%) are given in parentheses.}
 \label{fig:136ce-scheme}
 \end{center}
 \end{figure}

The isotope $^{136}$Ce is one of the most promising double beta
plus nuclei thanks to the high energy of the decay
($Q_{2\beta}=2378.55\pm0.27$ keV \cite{Wang:2017}) and a
comparatively high decay probability. However, the isotopic
abundance of $^{136}$Ce is rather low: $\delta=0.186 \pm 0.002$\%
\cite{Meija:2016}, which is typical for the double beta plus
active isotopes: their concentration in the natural isotopic
composition of elements is usually less than 1\%. A simplified
decay scheme of $^{136}$Ce is shown in Fig.
\ref{fig:136ce-scheme}. Due to the large decay energy $^{136}$Ce
is among the most studied nuclei. Double beta processes in the
cerium isotopes were searched for by using scintillation counting
\cite{Bernabei:1997,Danevich:2001,Belli:2003,Belli:2011} and
$\gamma$ spectrometric \cite{Belli:2009,Belli:2014} methods. In
the present study we use an ultra-low background HP Ge $\gamma$
spectrometer to search for $\gamma$ quanta of certain energies
expected in de-excitation of daughter nuclei, that is the
signature of the events under search.

\nopagebreak

\begin{figure}[htb]
\begin{center}
 \mbox{\epsfig{figure=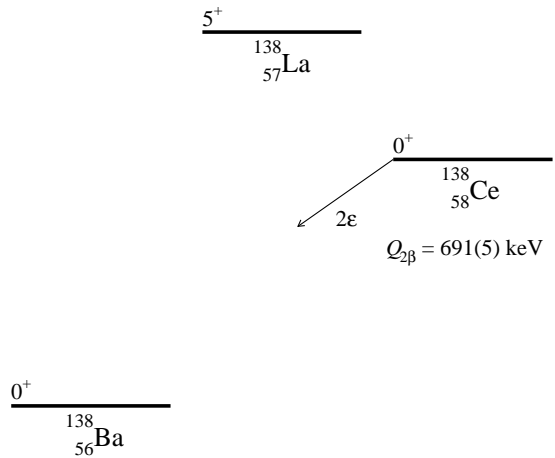,height=6.0cm}}
 \caption{Decay scheme of $^{138}$Ce.}
 \label{fig:138ce-scheme}
 \end{center}
 \end{figure}

Cerium contains two other potentially double beta active isotopes:
$^{138}$Ce ($Q_{2\beta}=691\pm5$ keV \cite{Wang:2017},
$\delta=0.251\pm0.002$\% \cite{Meija:2016}, the decay scheme is
presented in Fig. \ref{fig:138ce-scheme}) and $^{142}$Ce
($Q_{2\beta}=1416.8\pm2.2$ keV \cite{Wang:2017},
$\delta=11.114\pm0.051$\% \cite{Meija:2016}). The last isotope is
out of the present study since no $\gamma$ quanta are expected in
its $2\beta$ decay.

 \nopagebreak

In Section \ref{sec:pur} we describe the purification of the
cerium oxide sample. The experimental set-up and the radioactive
contamination of the sample after the purification will be
reported in Section \ref{sec:exp}. It should be stressed that deep
purification of cerium is also motivated in the light of radiopure
crystal scintillators development. Cerium can be used to develop
Ce-containing crystal scintillators (as for instance, CeF$_3$ and
CeCl$_3$), and as a dopant in inorganic scintillators:
Gd$_2$SiO$_5$(Ce), YAlO$_3$(Ce), LaBr$_3$(Ce), etc. The analysis
of the data, collected by using a sample of purified cerium oxide
in a HP Ge detector, and the obtained limits on $2\beta$ processes
in $^{136}$Ce and $^{138}$Ce are reported in Sect. \ref{sec:res}.

\section{Purification of cerium} \label{sec:pur}

Cerium oxide of 99\% TREO purity grade (CeO$_2$/TREO - 99.99\%)
was provided by the Stanford Materials Corporation and purified as
described in \cite{Belli:2014}. However, the concentration of
thorium remained rather high \cite{Belli:2014} (see also columns 3
and 4 of Table \ref{table:Ce-cont}). The contamination by thorium
(along with radium) has a major impact on $2\beta$ experiments
sensitivities. Thus, we decided to perform an additional
purification of the material (called 2nd purification in Table
\ref{table:Ce-cont}). The purification of the cerium oxide was
performed by using the liquid-liquid extraction method already
successfully applied for a variety of rare earths purification
from radioactive contamination \cite{Polischuk:2013}. The
purification has reduced the thorium concentration in the sample
by a factor of $\approx60$, that allowed to improve the
sensitivity of the experiment to the $2\varepsilon$,
$\varepsilon\beta^+$ and $2\beta^+$ decay of $^{136}$Ce by one
order of magnitude. The purification protocol is described in the
following.

\subsection{Cerium aqueous solution preparation}

Ultra-pure nitric and hydrofluoric acids solutions were used to
dissolve the CeO$_2$ powder to obtain a homogeneous aqueous
solution. The initial mixture ratios were calculated to achieve
both maximum cerium concentration and high acidity of the
solution. Very little additions of hydrofluoric acid and heating
of the system initiated the following process:

\begin{equation}
 \mbox{CeO}_2 + 6\mbox{HNO}_3 = \mbox{H}_2[\mbox{Ce(NO}_3)_6] +
 2\mbox{H}_2\mbox{O}.
\label{eg:1}
\end{equation}

At the same time two parallel reactions occur, which
decrease the concentrations of cerium and nitric acid:

\begin{equation}
 \mbox{Ce}^{4+}_{(aq.)}+4\mbox{F}^-_{(aq.)}=\mbox{CeF}_{4(solid)}\downarrow,
 \label{eg:2}
\end{equation}

\begin{equation}
 4\mbox{HNO}_{3(aq.)}=2\mbox{H}_2\mbox{O}_{(aq.)}+4\mbox{NO}_{2(gas)} \uparrow
 +~\mbox{O}_{2(gas)}\uparrow.
 \label{eg:3}
\end{equation}

The final solution composition was determined as 1.1
mol$\cdot$L$^{-1}$ of Ce$^{4+}$ and 11.7 mol$\cdot$L$^{-1}$ of
HNO$_3$.

\subsection{Preparation of organic extracting TBP liquid}

Tributyl phosphate (TBP, Acros Organics, 99+\%) has been chosen as
the most investigated solvent-extraction agent for lanthanides
separation. To decationize the organic liquid, 2
mol$\cdot$L$^{-1}$ nitric acid solutions were prepared from
ultra-pure HNO$_3$ and deionized water (18.2 M$\Omega \cdot $cm at
25 $^{\circ}$C) and then stirred for several minutes with an equal
volume of TBP. The residual aqueous solution was thrown away due to the
complete immiscibility of the phases, and the organic liquid was
rinsed with deionized water three times in a similar way.

\subsection{Extraction of cerium}

The extraction of cerium from the aqueous solution to the organic
phase of TBP can be described as:

\begin{equation}
 \mbox{H}_2[\mbox{Ce(NO}_{3})_{6}]_{(aq.)}+\mbox{nTBP}_{(org.)}=\mbox{H}_2[\mbox{Ce}\cdot
 \mbox{nTBP(NO}_{3})_{4}](\mbox{NO}_{3})_{2(org.)},~n=2-3.
 \label{eg:4}
\end{equation}

The competitive extraction of the cation impurities K$^+$,
Ra$^{2+}$, and lanthanides (Ln$^{3+}$) at the same conditions
should be very low, practically zero. However, Th$^{4+}$ has the
same chemical behavior as Ce$^{4+}$, thus it is extracted in a
similar way. Taking into account the initial cerium concentration
and stoichiometry in reaction (\ref{eg:4}), the liquid ratio was
taken 1$(aq.)$ to 0.75$(org.)$ on volume basis. The two liquids
were mixed together in a separation funnel, stirred for $2-3$
minutes and left for phase separation. After that, the organic
liquid was separated. The extraction degree has been determined to
be 84\%.

\subsection{Re-extraction of cerium}

After separation of the cerium contained organic solution in the
previous step, 0.1 M HNO$_3$ aqueous solution was used as
re-extracting liquid. Simultaneous reduction of Ce$^{4+}$ to
Ce$^{3+}$ by hydrogen peroxide addition was utilized to
decrease the distribution ratio between "organic" and "aqueous"
cerium:

\begin{multline}
 2\mbox{H}_2[\mbox{Ce}\cdot
 \mbox{nTBP(NO}_3)_4](\mbox{NO}_3)_{2(org)}+\mbox{H}_2\mbox{O}_{2(aq.)}=\\
 2\mbox{Ce(NO}_3)_{3(aq.)}+6\mbox{HNO}_{3(aq.)}+\mbox{O}_{2(gas)}+2\mbox{nTBP}_{(org.)}.
 \label{eg:5}
\end{multline}

\noindent The volumetric ratio of the two phases was 1 to 1. The
re-extraction of uranium is negligible at these conditions while
the removal of tetravalent thorium from organic to aqueous phase
can be explained by formation of an insoluble colloid thorium
peroxide compound. After the re-extraction procedure, the
obtained aqueous solution was heated for a while to decompose the
thorium peroxide and the hydrogen peroxide in excess. The determined
re-extraction degree of cerium was 65\% with respect to the initial
cerium oxide amount.

\subsection{Extraction of thorium}

Low acidic tri-$n$-octylphosphine oxide diluted to a concentration of 0.1
mol$\cdot$L$^{-1}$ in toluene has been used as
extraction agent for Th$^{4+}$ ions from the aqueous
solution containing cerium:

\begin{equation}
 \mbox{Ce}^{3+}_{(aq.)}+\mbox{Th}^{4+}_{(aq.)}+\mbox{nTOPO}_{(org.)}=\mbox{Ce}^{3+}_{(aq.)}+\mbox{Th}^{4+}\cdot
 \mbox{nTOPO}_{(org)}.
 \label{eg:6}
\end{equation}

\noindent The volumetric ratio was 1 of inorganic (aqueous) to 0.5
of organic phase. However, Ce$^{3+}$ does not leave the inorganic
phase in this process (\ref{eg:6}). After the separation of the
phases, the final purified cerium solution was prepared for the
last step of cerium oxide recovery.

\subsection{Cerium oxide recovery from solution}

The acidic cerium (III) nitrate solution was neutralized with
ammonia gas to reach a pH level up
to 7. The residual presence of hydrogen peroxide from the "cerium
re-extraction" step caused the chemical oxidation of Ce$^{3+}$ to
Ce$^{4+}$ in the low-acidic media that resulted in precipitation of
cerium hydroxyperoxide:

\begin{equation}
\mbox{2Ce}(\mbox{NO}_3)_3+\mbox{6NH}_3+\mbox{3H}_2\mbox{O}_2+\mbox{4H}_2\mbox{O}=\mbox{2Ce(OOH)(OH)}_3\downarrow+\mbox{6NH}_4\mbox{NO}_3.
\label{eg:7}
\end{equation}

The obtained amorphous sediment was decomposed by heating it
over $4-6$ hours at 90 $^{\circ}$C:

\begin{equation}
\mbox{2Ce(OOH)(OH)}_3=\mbox{2CeO(OH)}_2+\mbox{2H}_2\mbox{O}+\mbox{O}_2,
\label{eg:8}
\end{equation}

\noindent and separated from the liquid phase by using a centrifuge. The
final cerium oxide solid was prepared by annealing the CeO(OH)$_2$
at 900 $^{\circ}$C for 6 hours:

\begin{equation}
\mbox{CeO(OH)}_2=\mbox{CeO}_2+\mbox{H}_2\mbox{O}.
 \label{eg:9}
\end{equation}

As a result, 627 g of deeply purified CeO$_2$ were obtained
that is about 62\% of the initial amount of the material.

\section{Low counting experiment} \label{sec:exp}

The sample of deeply purified cerium oxide with mass 627 g,
encapsulated in a thin plastic container with internal sizes
$\oslash 90\times 50$ mm, was placed on the endcap of a high
purity germanium (HP Ge) detector named GeCris, with an active
volume of 465 cm$^3$, in the STELLA facility at the Laboratori
Nazionali del Gran Sasso. The detector is shielded by low
radioactive lead ($\approx25$ cm), copper ($\approx 5$ cm), and in
the innermost part by Roman lead ($\approx 2.5$ cm). The set-up is
enclosed in an air-tight PMMA box and permanently flushed with
High Purity boiling-off nitrogen to reduce the radon
concentration. The energy resolution of the detector (full width
at half maximum, FWHM) depends on the energy $E_{\gamma}$ of the
$\gamma$ quanta as FWHM(keV)~$=\sqrt{1.41+0.00197\times
E_{\gamma}}$, where $E_{\gamma}$ is in keV.

\begin{figure}[htb]
\begin{center}
 \mbox{\epsfig{figure=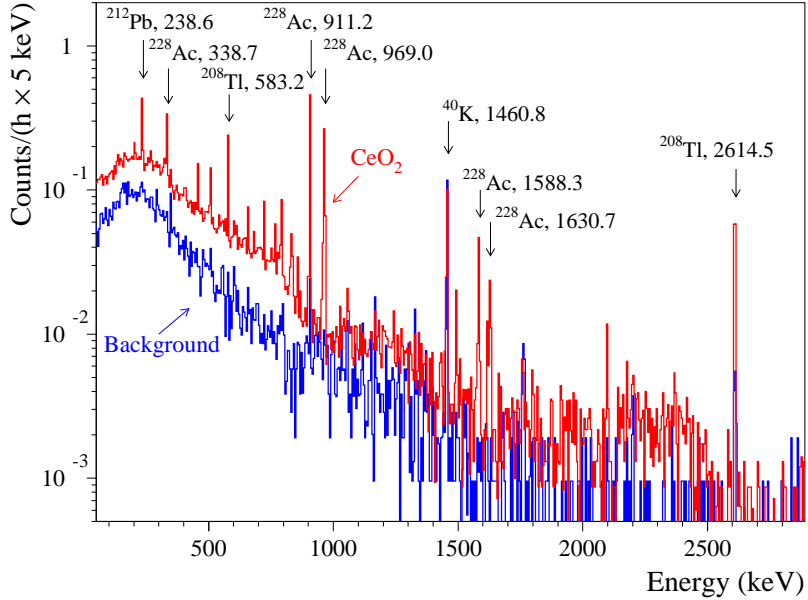,height=8.0cm}}
\caption{Energy spectra measured by the GeCris HP Ge detector with
cerium oxide sample (CeO$_2$) over 2299 h and background
accumulated over 1046 h (Background). The counting rate in the
energy spectrum accumulated with the  CeO$_2$ sample is
substantially lower in comparison to the previous experiment (see
Fig. 3 in \cite{Belli:2014}) thanks to the deep purification of
the material described in Section \ref{sec:pur}. Some excess in
the CeO$_2$ data is mainly due to the residual contamination of
the sample by radium ($^{228}$Ra). Energy of $\gamma$ quanta are
in keV.}
 \label{fig:BG}
 \end{center}
 \end{figure}

The energy spectrum measured by the HP Ge detector with the
CeO$_2$ sample over 2299 h is presented in Fig. \ref{fig:BG}
together with background data accumulated over 1046 h. The
counting rate in the energy spectrum measured with the CeO$_2$
sample is substantially lower than that after the first stage of
the material purification \cite{Belli:2014}. There is some excess
in the CeO$_2$ data due to the residual contamination of the
sample, mainly by radium ($^{228}$Ra). Activities of radionuclides
in the CeO$_2$ sample ($A$) were estimated by using the following
formula:

\begin{equation}
A = (S_{sample}/t_{sample} - S_{bg}/t_{bg})/(\eta_{\gamma} \cdot
m),
\end{equation}

\noindent where $S_{sample}$ ($S_{bg}$) is the area of a peak in
the sample (background) spectrum; $t_{sample}$ ($t_{bg}$) is the
time of the sample (background) measurement; $\eta_{\gamma}$ is
the detection efficiency of $\gamma$ quanta in the full absorbtion
peak; $m$ is the mass of the sample. The detection efficiencies
were estimated by Monte Carlo simulations with EGSnrc \cite{EGS}
and GEANT4 \cite{GEANT4} packages as well as the event generator
DECAY0 \cite{DECAY0,DECAY0-prep}. Both the codes give compatible
values of the detection efficiencies for $\gamma$ quanta of
$^{40}$K, $^{137}$Cs, $^{138}$La, $^{139}$Ce, $^{152}$Eu,
$^{154}$Eu, $^{176}$Lu and $\gamma$-emitting daughters of
$^{232}$Th, $^{235}$U and $^{238}$U. The efficiencies calculated
by GEANT4 and EGSnrs differ on average in $\sim3\%$, with a
maximal difference 12.7\% (for the low intensity $\gamma$ quanta
1001.0 keV of $^{234m}$Pa from the $^{238}$U chain). The EGSnrs
code systematically gives slightly bigger efficiencies. The
statistical errors of the Monte Carlo calculations are within
$0.5\%-0.6\%$. We have used weighted mean values of the detection
efficiencies obtained by the two codes to calculate the activities
(or upper limits) of contaminants in the CeO$_2$ sample, since we
have not found arguments to prefer one of the codes. The detection
efficiencies include the branching ratios for the $\gamma$ quanta
incorporated in the DECAY0 event generator
\cite{DECAY0,DECAY0-prep}. The upper limits were set in the cases
when the values and errors of the difference $(S_{sample} -
S_{bg}\cdot t_{sample}/t_{bg})$ give no signature of the $\gamma$
peak observation. The results of the CeO$_2$ sample radiopurity
analysis are presented in Table \ref{table:Ce-cont}.

\nopagebreak

\begin{table}[htb]
\caption{Radioactive contamination of the cerium oxide before and
after two stages of purification. The reference date for the
activities of the CeO$_2$ sample after the 2nd purification is
April 12, 2016. Upper limits are given at 90\% C.L., the
uncertainties of the measured activities are given at 68\% C.L.}
\begin{center}
\begin{tabular}{lllll}
\hline
 Chain      & Nuclide     & \multicolumn{3}{c}{Activity (mBq kg$^{-1}$)} \\
\cline{3-5}
~           & ~                     & before        & after 1st                         & after 2nd    \\
~           & ~                     & purification  \cite{Belli:2014} & purification \cite{Belli:2014}    & purification \\
\hline
 ~          & $^{40}$K              & 77(28)        & $\leq9$                           & $\leq 4$ \\
 ~          & $^{137}$Cs            & $\leq 3$      & $\leq2$                           & $0.4\pm0.2$       \\
 ~          & $^{138}$La            & --            & $\leq0.7$                         & $\leq 0.6$    \\
 ~          & $^{139}$Ce            & --            & $6\pm1$                         & $1.4\pm0.3$    \\
 ~          & $^{152}$Eu            & --            & $\leq0.5$                         & $\leq 0.2$       \\
 ~          & $^{154}$Eu            & --            & $\leq0.9$                         & $\leq 0.08$     \\
 ~          & $^{176}$Lu            & --            & $\leq0.5$                         & $0.4\pm 0.1$   \\

 $^{232}$Th & $^{228}$Ra            & $850\pm50$    & $53\pm3$                          & $30.4 \pm 0.7$   \\
 ~          & $^{228}$Th            & $620\pm30$    & $573\pm17$                        & $9.8 \pm 0.5$   \\
 ~ & ~ & ~ & ~ & ~\\
 $^{235}$U  & $^{235}$U             & $38\pm10$     & $\leq1.8$                         & $\leq 0.4$ \\
 ~          & $^{231}$Pa            & --            & $\leq24$                          & $\leq 0.4$   \\
 ~          & $^{227}$Ac            & --            & $\leq3$                           & $\leq 1.4$    \\
 ~ & ~ & ~ & ~ & ~\\
 $^{238}$U  & $^{238}$U             & $\leq870$     & $\leq40$                          & $\leq 12$   \\
 ~          & $^{226}$Ra            & $11\pm3$      & $\leq1.5$                         & $\leq 0.3$   \\
 ~ & ~ & ~ & ~ & ~\\
\hline
\end{tabular}
  \label{table:Ce-cont}
\end{center}
\end{table}

\section{Results for $2\beta$ processes and discussion}
\label{sec:res}

Gamma quanta of certain energies are expected in de-excitation of
daughter nuclei in the $2\beta$ processes in $^{136}$Ce and
$^{138}$Ce according to the decay schemes (Figs.
\ref{fig:136ce-scheme} and \ref{fig:138ce-scheme}). We suppose
that, in result of $^{136}$Ce decay, some excited level (or the
ground state) of $^{136}$Ba is populated with 100\% probability.
In the subsequent de-excitation process, probabilities of
transitions to different lower levels with emission of $\gamma$
quanta (or conversion electrons or $e^+e^-$  pairs) are taken from
\cite{Sonzogni:2002}. All these probabilities are incorporated in
the DECAY0 event generator \cite{DECAY0,DECAY0-prep}. In
neutrinoless $2\varepsilon$ process, the emission of one   quantum
with energy $E_{\gamma} = Q_{2\beta}-E_{exc}-E_{b_1}-E_{b_2}$ is
supposed, where $E_{exc}$ is the energy of the excited level, and
$E_{b_i}$ is the binding energy of the i-th captured electron on
the atomic shell.

Since there are no peculiarities in the spectrum of the CeO$_2$
sample, which could be ascribed to the $2\beta$ decay of
$^{136}$Ce or $^{138}$Ce, we have calculated lower half-life
limits using the following formula:

\begin{equation}
\lim T_{1/2} = N \cdot \eta \cdot t \cdot \ln 2 / \lim S,
\end{equation}

\noindent where $N$ is the number of $^{136}$Ce or $^{138}$Ce
nuclei in the CeO$_2$ sample, $\eta$ is the detection efficiency,
$t$ is the measuring time, and $\lim S$ is the number of events of
the effect searched for which can be excluded at a given
confidence level (C.L.). The CeO$_2$ sample contains
$4.06\times10^{21}$ and $5.51\times10^{21}$ nuclei of $^{136}$Ce
and $^{138}$Ce, respectively. The full absorbtion peak detection
efficiencies of the set-up to the $\gamma$ quanta expected in the
double beta decay of $^{136}$Ce and $^{138}$Ce were calculated
using the EGSnrc and GEANT4 packages together with the event
generator DECAY0. The efficiencies calculated by GEANT4 and EGSnrs
differ on average in $\sim 1.5\%$ with a maximal difference
$5.7\%$ (the EGSnrs code again gives higher efficiencies). The
statistical errors of all the calculations are within
$0.3\%-0.6\%$. Thus, as in the contamination analysis, we have
used weighted mean values of the detection efficiencies. The
detection efficiencies are given in Table
\ref{table:Ce_2b_limits}.

The sensitivity of the experimental set-up to the
$2\nu2\varepsilon$ decay of $^{136}$Ce and $^{138}$Ce is almost
negligible due to the very low detection efficiency of the detector
for the X rays expected in the decays. Thus, we omit estimations on
these decay half-lives.

\begin{figure}[htb]
\begin{center}
 \mbox{\epsfig{figure=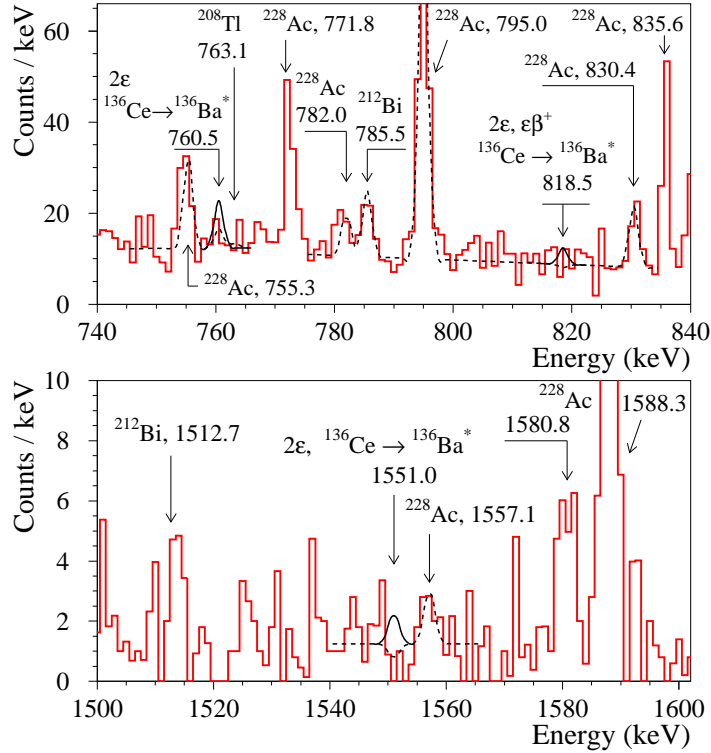,height=10.0cm}}
\caption{Energy spectrum measured with cerium oxide sample
(CeO$_2$) over 2299 h in the energy intervals where $\gamma$ peaks
with energies 760.5 keV, 818.5 keV and 1551.0 keV are expected.
The excluded peaks are shown by solid lines, the backgrounds for
the processes are shown by dashed lines.}
 \label{fig:818&1551}
 \end{center}
 \end{figure}

In the case of double electron capture in $^{136}$Ce to the
excited states of $^{136}$Ba, $\gamma$ quanta can be detected with
reasonable efficiencies at a level of a few \%. The parts of
energy spectrum collected with the CeO$_2$ sample in the vicinity
of the expected $\gamma$ quanta 760.5 keV, 818.5 keV and 1551.0
keV of the $2\varepsilon$ decays of $^{136}$Ce to the excited
states of $^{136}$Ba are shown in Fig. \ref{fig:818&1551}. A fit
of the data in the energy interval $745-767$ keV was done by using
a model consisting of three Gaussian functions and a polynomial
function of the 1st degree is used to express the continuous
background. Two Gaussian functions describe the $\gamma$ peaks in
the vicinity of the energy of interest: 755.3 keV ($^{228}$Ac) and
763.1 keV ($^{208}$Tl), while the third one with energy 760.5 keV
describes the peak expected in the $2\varepsilon$ decay of
$^{136}$Ce to the $0^+$ 1579 keV excited level of $^{136}$Ba (the
fit function is shown in Fig. \ref{fig:818&1551}). The positions
and energy resolutions of the peaks were fixed according to the
expected energies and the measured dependence of the HP Ge
detector energy resolution on energy of $\gamma$ quanta. The fit
gives an area of the 760.5 keV peak $S=8.0 \pm 6.9$ counts (the
quality of the fit is characterized by the value of
$\chi^2$/n.d.f. = 1.55), that is no evidence on the effect
searched for. An upper limit $\lim S=19.3$ counts was obtained by
using the Feldman-Cousins procedure \cite{Fel98}.

However, taking into account a more favorable background condition
in the vicinity of the expected 818.5 keV peak, and almost the
same detection efficiency ($\eta=2.31\%$ for 760.5 keV and
$\eta=2.60\%$ for 818.5 keV $\gamma$ quanta), the analysis of
818.5 keV line is more sensitive to the $2\varepsilon$ decays of
$^{136}$Ce to the $0^+$ 1579 keV excited state of $^{136}$Ba. We
have used the following model of background in the energy interval
$775-835$ keV: a polynomial function of the 1st degree to describe
the continuous background, $\gamma$ peaks at energies 782.0 keV,
785.5 keV, 795.0 keV, and 830.4 keV ($\gamma$ quanta of $^{228}$Ac
and $^{212}$Bi, see Fig. \ref{fig:818&1551}) and the 818.5 keV
peak searched for. We should use such a complicated model to
describe the background in the vicinity of the energy 818.5 keV
correctly. The fit gives area of the peak $S = -1.5 \pm 4.9$
counts ($\chi^2$/n.d.f. = 1.47). Since this value gives no
evidence for the effect searched for, we have estimated the number
of events which can be excluded at 90\% C.L. as $\lim S = 6.6$
counts using the approach given in \cite{Fel98}. The excluded peak
is shown in Fig. \ref{fig:818&1551}. Thus, the lower limit could
be increased by almost one order of magnitude with respect to the
previous experiment \cite{Belli:2014}. The new lower bound on the
half-life of the two neutrino double electron capture in
$^{136}$Ce to the $0^+$ 1579.0 keV excited level of $^{136}$Ba is
$T_{1/2}\geq 2.5\times10^{18}$ yr. The $\lim S$ estimation was
used to set limits also for decays to the first $2^+$ 818.5 keV,
and some other $2\varepsilon$ and $\varepsilon\beta^+$ transitions
in $^{136}$Ce towards excited levels of $^{136}$Ba (see Table
\ref{table:Ce_2b_limits}).

To set a limit on the 1551.0 keV peak we added to the background
model also the $\gamma$ peak of $^{228}$Ac with energy 1557.1 keV.
The result of the fit is shown in the lower panel of Fig.
\ref{fig:818&1551}. Limits on other double electron capture
processes in $^{136}$Ce to the excited levels of $^{136}$Ba,
presented in Table \ref{table:Ce_2b_limits}, were set in a similar
way.

\nopagebreak

\begin{figure}[htb]
\begin{center}
 \mbox{\epsfig{figure=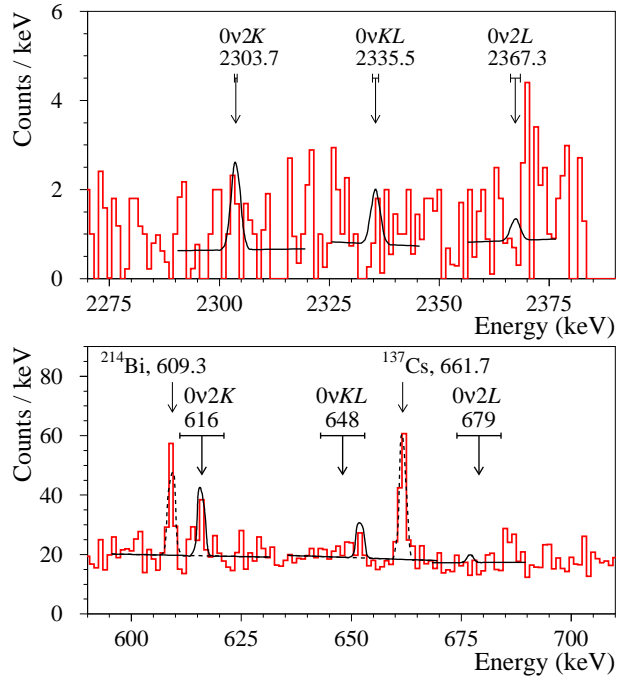,height=9.0cm}}
\caption{Parts of the energy spectrum collected with the CeO$_2$
sample where peaks from $0\nu2\varepsilon$ processes ($2K$, $KL$
and $2L$) in $^{136}$Ce (upper panel) and $^{138}$Ce (lower panel)
are expected. The excluded peaks for the processes are shown by
solid lines (the peculiarities providing a maximal area of the
expected peak within the interval of interest were considered).
The peaks of $^{214}$Bi with energy 609.3 keV and of $^{137}$Cs
with energy 661.7 keV are shown by dashed lines.}
 \label{fig:0n2e}
 \end{center}
 \end{figure}

In the case of the $0\nu$ double electron capture in $^{136}$Ce
and $^{138}$Ce from $K$ and $L$ shells to the ground states of the
daughter nuclei, the energies of the $\gamma$ quanta are expected
to be equal to $E_{\gamma}=Q_{2\beta}-E_{b_1}-E_{b_2}$, where
$E_{b_i}$ are the binding energies of the captured electrons on
the $K$ and $L$ atomic shells of the daughter nuclei. The energy
spectra accumulated with the CeO$_2$ sample in the vicinity of the
expected energies of the quanta for the $0\nu2\varepsilon$ capture
in $^{136,138}$Ce to the ground state of $^{136,138}$Ba are shown
in Fig. \ref{fig:0n2e}. In the case of $^{138}$Ce
$0\nu2\varepsilon$ decay the energies of the $\gamma$ quanta are
known with an accuracy of $\pm 5$ keV. Therefore, we had to
consider larger energy intervals of interest to estimate the
limits on the $0\nu$ double electron capture in $^{136,138}$Ce
from $K$ and $L$ shells. In the case of $^{138}$Ce also the peaks
of $^{214}$Bi with energy 609.3 keV, and of $^{137}$Cs with energy
661.7 keV had to be taken into account to build a correct
background model in a wide enough energy interval near the
expected peaks. The obtained upper limits are given in Table
\ref{table:Ce_2b_limits}.

\nopagebreak

\begin{figure}[htb]
\begin{center}
 \mbox{\epsfig{figure=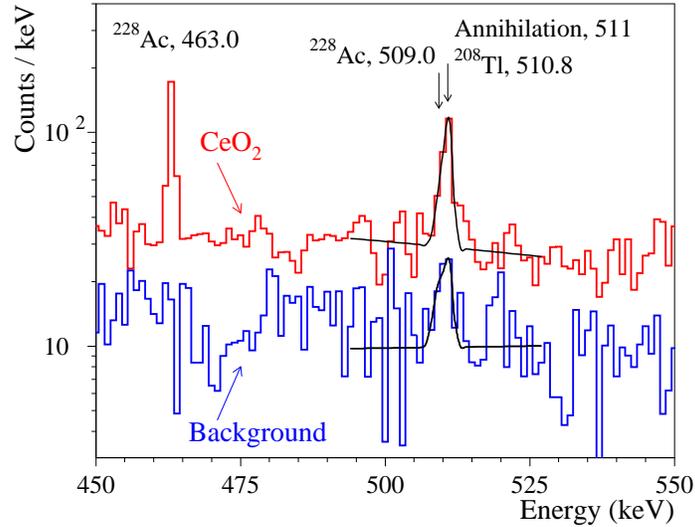,height=7.0cm}}
\caption{Energy spectrum measured with the cerium oxide sample
(CeO$_2$) over 2299 h and background (1046 h, Background) in the
vicinity of the 511 keV annihilation peak.}
 \label{fig:511}
 \end{center}
 \end{figure}

The electron capture with positron and double positron emission in
$^{136}$Ce should lead to the emission of annihilation $\gamma$
quanta with energy 511 keV. Thus the bounds on the processes were
set by analysing the annihilation $\gamma$ peak in the data taking
into account the 511 keV peak in the background and the
contributions in this peak from the contaminants present in the
sample. The energy spectrum measured with the CeO$_2$ sample and
the background data in the vicinity of the annihilation peak are
shown in Fig. \ref{fig:511}. Taking into account the radioactive
contamination of the CeO$_2$ by $^{228}$Ra and $^{228}$Th we have
included in the background model the 509.0 keV peak of $^{228}$Ac
and 510.8 keV peak of $^{208}$Tl. The area of the
$509.0+510.8+511$ keV peak in the CeO$_2$ spectrum is $182\pm26$
counts, while the peak in the background data has an area of
$20\pm7$ counts ($44\pm14$ counts taking into account the
different measurements live times). According to the Monte Carlo
simulations the radioactivity of $^{228}$Ac in the sample provides
$33\pm1$ counts in the sum peak, and $^{228}$Th contributes
$155\pm8$ counts to the annihilation peak. Thus, the difference is
$-50\pm31$ counts, which corresponds (according to \cite{Fel98})
to $\lim S=16$ counts at 90\% C.L. The estimation leads to the new
improved limits on the $\varepsilon\beta^+$ and $2\beta^+$ decays
of $^{136}$Ce. Again, all the obtained limits on the double beta
decay processes in $^{136}$Ce and $^{138}$Ce are summarized in
Table \ref{table:Ce_2b_limits}.

\nopagebreak

\clearpage

\begin{landscape}

\begin{table*}[htbp]
\caption{The half-life limits on 2$\beta$ processes in $^{136}$Ce
and $^{138}$Ce together with the best previous limits and
theoretical predictions (the theoretical $T_{1/2}$ values for
$0\nu$ mode are given for $m_{\nu}$ = 1 eV). The energies of the
$\gamma$ lines ($E_{\gamma}$), which were used to set the
$T_{1/2}$ limits, are listed with their corresponding detection
efficiencies ($\eta$) and values of $\lim S$.}
\begin{center}
\resizebox{1.4\textwidth}{!}{
\begin{tabular}{lllllllll}

\hline
 Process  & Decay   & Level of & $E_{\gamma}$  & $\eta$  & $\lim S$ & \multicolumn{2}{c}{Experimental limits,}         & Theoretical estimations, \\
 of decay & mode    & daughter & (keV)         & (\%)    & (cnt)    &  \multicolumn{2}{c}{$T_{1/2}$ (yr) at 90\% C.L.} & $T_{1/2}$ (yr)  \\
\cline{7-8}
 ~        & ~       & nucleus  & ~             & ~       & at 90\%  & Present work            & Best previous       & ~                \\
 ~        & ~       & (keV)    & ~             & ~       & C.L.  & ~                       & results             & ~    \\
 \hline

\multicolumn{9}{l}{$^{136}$Ce $\to$ $^{136}$Ba} \\

 $2\varepsilon$       & 2$\nu$ & $2^+$ 818.5  & 818.5  & 2.60 & 6.6 & $\geq2.9\times10^{18}$ & $\geq3.3\times10^{17}$ \cite{Belli:2014} & -- \\
 ~                    & ~      & $2^+$ 1551.0 & 1551.0 & 1.07 & 2.3 & $\geq3.4\times10^{18}$ & $\geq6.9\times10^{17}$ \cite{Belli:2014} & -- \\
 ~                    & ~      & $0^+$ 1579.0 & 818.5  & 2.26 & 6.6 & $\geq2.5\times10^{18}$ & $\geq1.6\times10^{18}$ \cite{Belli:2014} & -- \\
 ~                    & ~      & $2^+$ 2080.0 & 2080.0 & 0.624 & 1.64 & $\geq2.8\times10^{18}$ & $\geq1.7\times10^{18}$ \cite{Belli:2014} & -- \\
 ~                    & ~      & $2^+$ 2128.8 & 2128.8 & 0.589 & 3.1 & $\geq1.4\times10^{18}$ & $\geq9.1\times10^{17}$ \cite{Belli:2014} & -- \\
 ~                    & ~      & $0^+$ 2141.3 & 1322.8 & 1.84  & 2.4 & $\geq4.4\times10^{18}$ & $\geq2.9\times10^{17}$ \cite{Belli:2014} & -- \\
 ~                    & ~      & $(2)^+$ 2222.7 & 671.7 & 1.09 & 4.1 & $\geq2.0\times10^{18}$ & $\geq7.4\times10^{17}$ \cite{Belli:2014} & -- \\
 ~                    & ~      & $0^+$ 2315.3 & 818.5 & 2.26 & 6.6 & $\geq2.5\times10^{18}$ & $\geq2.9\times10^{17}$ \cite{Belli:2014} & -- \\

 2$K$                 & 0$\nu$ & g.s. & 2303.7 & 1.50 & 5.2 & $\geq2.1\times10^{18}$ & $\geq4.6\times10^{17}$ \cite{Belli:2014} & -- \\
 $KL$                 & 0$\nu$ & g.s. & 2335.5 & 1.49 & 3.2 & $\geq3.4\times10^{18}$ & $\geq6.6\times10^{17}$ \cite{Belli:2014} & -- \\
 2$L$                 & 0$\nu$ & g.s. & 2367.3 & 1.48 & 1.3 & $\geq8.4\times10^{18}$ & $\geq5.4\times10^{17}$ \cite{Belli:2014} & -- \\

 2$\varepsilon$       & 0$\nu$ & $2^+$ 818.5  & 1485.5 & 1.73 & 4.2 & $\geq3.0\times10^{18}$ & $\geq2.5\times10^{18}$ \cite{Belli:2014} & -- \\
 ~                    & ~      & $2^+$ 1551.0 & 1551.0 & 0.914 & 2.3 & $\geq2.9\times10^{18}$ & $\geq6.0\times10^{17}$ \cite{Belli:2014} & -- \\
 ~                    & ~      & $0^+$ 1579.0 & 818.5  & 1.95 & 6.6 & $\geq2.2\times10^{18}$ & $\geq1.1\times10^{18}$ \cite{Belli:2014} & -- \\
 ~                    & ~      & $2^+$ 2080.0 & 2080.0 & 0.580 & 1.64 & $\geq2.6\times10^{18}$ & $\geq1.6\times10^{18}$ \cite{Belli:2014} & -- \\
 ~                    & ~      & $2^+$ 2128.8 & 818.5 & 1.43 & 6.6 & $\geq1.6\times10^{18}$ & $\geq8.6\times10^{17}$ \cite{Belli:2014} & -- \\
 ~                    & ~      & $0^+$ 2141.3 & 1322.8 & 1.76 & 3.12 & $\geq4.2\times10^{18}$ & $\geq2.8\times10^{17}$ \cite{Belli:2014} & -- \\
 ~                    & ~      & $(2)^+$ 2222.7 & 671.7 & 1.09 & 4.10 & $\geq2.0\times10^{18}$ & $\geq7.3\times10^{17}$ \cite{Belli:2014} & -- \\
 ~                    & ~      & $0^+$ 2315.3 & 818.5 & 2.26 & 6.6 & $\geq2.5\times10^{18}$ & $\geq2.9\times10^{17}$ \cite{Belli:2014} & $1.0\times10^{23}-2.3\times10^{33}$ \cite{Kolhinen:2011,Krivoruchenko:2011,Suhonen:2012} \\

 $\varepsilon\beta^+$ & 2$\nu$ & g.s. & 511 & 5.90 & 16 & $\geq1.0\times10^{17}$ & $\geq2.7\times10^{18}$ \cite{Belli:2014} & (6.0--28)$\times10^{23}$ \cite{Hirsch:1994,Rumyantsev:1998,Abad:1984} \\
 ~                    & ~      & $2^+$ 818.5  & 511  & 5.16 & 16 & $\geq2.4\times10^{18}$ & $\geq2.5\times10^{17}$ \cite{Belli:2014} & -- \\
 ~                    & 0$\nu$ & g.s. & 511 & 5.68 & 16 & $\geq2.6\times10^{18}$ & $\geq9.6\times10^{16}$ \cite{Belli:2014} & $(2.7-4.7)\times10^{26}$ \cite{Hirsch:1994,Suhonen:2003,Barea:2013} \\
 ~                    & ~      & $2^+$ 818.5  & 511  & 5.09 & 16 & $\geq2.3\times10^{18}$ & $\geq2.5\times10^{17}$ \cite{Belli:2014} & -- \\

 2$\beta^+$ & 2$\nu$ & g.s. & 511 & 8.85 & 16 & $\geq4.1\times10^{18}$ & $\geq3.5\times10^{17}$ \cite{Belli:2014} & $(5.2-9.6)\times10^{31}$ \cite{Hirsch:1994,Abad:1984} \\
 ~          & 0$\nu$ & g.s. & 511 & 8.80 & 16 & $\geq4.1\times10^{18}$ & $\geq6.9\times10^{17}$ \cite{Bernabei:1997} & (1.7--2.7)$\times10^{29}$  \cite{Hirsch:1994,Suhonen:2003,Barea:2013} \\

 \hline
 \multicolumn{9}{l}{$^{138}$Ce $\to$ $^{138}$Ba} \\

 2$K$                 & 0$\nu$ & g.s. & $616\pm5$ & 2.86 & 45 & $\geq4.7\times10^{17}$ & $\geq5.5\times10^{17}$ \cite{Belli:2014} & \\ 
 $KL$                 & 0$\nu$ & g.s. & $648\pm5$ & 2.82 & 25 & $\geq8.3\times10^{17}$ & $\geq4.6\times10^{17}$ \cite{Belli:2014} & -- \\
 2$L$                 & 0$\nu$ & g.s. & $679\pm5$ & 2.77 & 4.9 & $\geq4.2\times10^{18}$ & $\geq4.0\times10^{17}$ \cite{Belli:2014} & -- \\
 \hline

 \end{tabular}
 }
 \end{center}
 \label{table:Ce_2b_limits}
 \end{table*}

\end{landscape}

\clearpage

\section{Conclusions}

Deep purification of cerium oxide from radioactive contamination
was obtained using the liquid-liquid extraction method. Measurements performed by
ultra-low background $\gamma$-ray spectrometry have demonstrated a
high efficiency of the purification method. The radiopurity of the CeO$_2$
sample is at the level of (less than) mBq kg$^{-1}$ -- tenths of mBq kg$^{-1}$ of
$^{137}$Cs, $^{138}$La, $^{139}$Ce (cosmogenic radionuclide),
$^{152}$Eu, $^{154}$Eu, $^{176}$Lu, $^{226}$Ra and
$\gamma$-emitting daughters of $^{235}$U. The activity of
$^{238}$U (estimated via $^{234m}$Pa) does not exceed 12 mBq kg$^{-1}$,
while $^{228}$Ra and $^{228}$Th remain in the sample at the level
of 30 mBq kg$^{-1}$ and 10 mBq kg$^{-1}$, respectively. Thus, the thorium
concentration in the sample was reduced with this
purification by a factor of 60. The purification protocol can be
applied to produce radiopure cerium-containing crystal
scintillators for the next stage experiments to search for double
beta decay of cerium.

The experiment sets the best up-to-date limits on different modes
and channels of double beta decay of $^{136}$Ce and $^{138}$Ce at
the level of $T_{1/2}>10^{17}-10^{18}$ yr. The sensitivity of the
experiment is still far from the theoretical predictions (given in
Table \ref{table:Ce_2b_limits}). Further improvement of the
experimental sensitivity can be achieved by using enriched
materials, by further reduction of the background, and by the
application of cerium-containing crystal scintillators to increase
the detection efficiency.


\begin{thebibliography}{99}

 \bibitem{Barea:2012} J.~Barea, J.~Kotila, F.~Iachello, Limits on Neutrino Masses from Neutrinoless Double-$\beta$ Decay, Phys. Rev. Lett. 109 (2012) 042501.

 \bibitem{Rodejohann:2012} W.~Rodejohann, Neutrino-less double $\beta$ decay and particle physics, J. Phys. G 39 (2012) 124008.

 \bibitem{Delloro:2016} S.~Dell'Oro, S.~Marcocci, M.~Viel, F.~Vissani, Neutrinoless Double Beta Decay: 2015 Review, AHEP 2016 (2016) 2162659.

 \bibitem{Vergados:2016} J.D. Vergados, H. Ejiri, F. Simkovic, Neutrinoless double beta decay and neutrino mass, Int. J. Mod. Phys. E 25 (2016) 1630007.

 \bibitem{Deppisch:2012} F.F.~Deppisch, M.~Hirsch, H.~P\"{a}s, Neutrinoless double-$\beta$ decay and physics beyond the standard model, J. Phys. G 39 (2012) 124007.

 \bibitem{Bilenky:2015} S.M.~Bilenky, C.~Giunti, Neutrinoless double-$\beta$ decay: A probe of physics beyond the standard model, Int. J. Mod. Phys. A 30 (2015) 1530001.

\bibitem{Tretyak:2002} V.I.~Tretyak, Yu.G.~Zdesenko, Tables of double $\beta$ decay data -- an update, At. Data Nucl. Data Tables 80 (2002) 83.

\bibitem{Elliott:2012} S.R.~Elliott, Recent progress in double beta decay, Mod. Phys. Lett. A 27 (2012) 123009.

\bibitem{Giuliani:2012} A.~Giuliani, A.~Poves, Neutrinoless Double-Beta Decay, AHEP 2012 (2012) 857016.

\bibitem{Cremonesi:2014} O.~Cremonesi, M.~Pavan, Challenges in Double Beta Decay, AHEP 2014 (2014) 951432.

\bibitem{Gomes:2015} J.J.~G$\mathrm{\acute{o}}$mez-Cadenas, J.~Mart$\mathrm{\acute{i}}$n-Albo, Phenomenology of Neutrinoless Double Beta Decay, Proc. of Sci. (GSSI14) 004 (2015).

\bibitem{Sarazin:2015} X.~Sarazin, Review of Double Beta Experiments, J. Phys.: Conf. Ser. 593 (2015) 012006.

\bibitem{GERDA} M.~Agostini et al., Background-free search for neutrinoless double-$\beta$ decay of $^{76}$Ge with GERDA, Nature 544 (2017) 47.

\bibitem{EXO-200} J.B.~Albert et al. (The EXO-200 Collaboration), Search for Majorana neutrinos with the first two years of EXO-200
data, Nature 510 (2014) 229.

\bibitem{CUORE} K. Alfonso et al. (CUORE Collaboration), Search for Neutrinoless Double-Beta Decay of $^{130}$Te with
CUORE-0, Phys. Rev. Lett. 115 (2015) 102502.

\bibitem{NEMO-3} R. Arnold et al., Results of the search for neutrinoless double-$\beta$ decay in $^{100}$Mo with the NEMO-3
experiment, Phys. Rev. D 92 (2015) 072011.

\bibitem{KamLAND-Zen} A.~Gando et al. (KamLAND-Zen Collaboration), Search for Majorana Neutrinos Near the Inverted Mass Hierarchy Region with KamLAND-Zen, Phys. Rev. Lett.
117 (2017) 082503.

 \bibitem{Maalampi:2013} J.~Maalampi, J.~Suhonen, Neutrinoless Double $\beta^+$/EC Decays, AHEP 2013 (2013) 505874.

 \bibitem{Hirsch:1994} M.~Hirsch, K.~Muto, T.~Oda, H.V.~Klapdor-Kleingrothaus, Nuclear structure calculation of $\beta^+ \beta^+$, $\beta^+$/EC and EC/EC decay matrix elements, Z. Phys. A 347 (1994) 151.

 \bibitem{Bernabeu:1983} J.~Bernabeu, A.~De Rujula, C.~Jarlskog, Neutrinoless double electron capture as a tool to measure the electron neutrino mass, Nucl. Phys. B 223 (1983) 15.

 \bibitem{Kolhinen:2011} V.S.~Kolhinen et al., On the resonant neutrinoless double-electron-capture decay of $^{136}$Ce, Phys. Lett. B 697 (2011) 116.

 \bibitem{Kotila:2014} J.~Kotila, J.~Barea, F.~Iachello, Neutrinoless double-electron capture, Phys. Rev. C 89 (2014) 064319.

 \bibitem{Doi:1993} M.~Doi, T.~Kotani, Neutrinoless modes of double beta decay, Prog. Theor. Phys. 89 (1993) 139.

 \bibitem{Wang:2017} M.~Wang et al., The AME2016 atomic mass evaluation, (II). Tables, graphs and references, Chin. Phys. C 41 (2017) 030003.

 \bibitem{Meija:2016} J.~Meija et al., Isotopic compositions of the elements 2013 (IUPAC Technical Report), Pure Appl. Chem. 88 (2016) 293.

 \bibitem{Bernabei:1997} R.~Bernabei et al., Feasibility of $\beta \beta$ decay searches with Ce isotopes using CeF$_3$ scintillators, Nuovo Cimento A 110 (1997) 189.
 \bibitem{Danevich:2001} F.A.~Danevich et al., Quest for double beta decay of $^{160}$Gd and Ce isotopes, Nucl. Phys. A 694 (2001) 375.
 \bibitem{Belli:2003} P.~Belli et al., Performances of a CeF$_3$ crystal scintillator and its application to the search for rare processes, Nucl. Instrum. Meth.  A 498 (2003) 352.
 \bibitem{Belli:2011} P.~Belli et al., Search for $2\beta$ decay of cerium isotopes with CeCl$_3$ scintillator, J. Phys. G 38 (2011) 015103.
 \bibitem{Belli:2009} P.~Belli et al., First limits on neutrinoless resonant $2\varepsilon$ captures in $^{136}$Ce and new limits for other $2\beta$ processes in $^{136}$Ce and $^{138}$Ce isotopes, Nucl. Phys. A 824 (2009) 101.
 \bibitem{Belli:2014} P.~Belli et al., Search for double beta decay of $^{136}$Ce and $^{138}$Ce with HPGe gamma detector, Nucl. Phys. A 930 (2014) 195.

 \bibitem{Polischuk:2013} O.G.~Polischuk et al., Purification of lanthanides for double beta decay
 experiments, AIP Conf. Proc. 1549 (2013) 124.

 \bibitem{EGS} I.~Kawrakow, D.W.O.~Rogers, The EGSnrc code system: Monte Carlo simulation of electron and photon transport, NRCC Report PIRS-701, Ottawa, 2003.
 \bibitem{GEANT4} S.~Agostinelli et al., GEANT4 -- a simulation toolkit, Nucl. Instrum. Meth. A 506 (2003) 250.
 \bibitem{DECAY0} O.A.~Ponkratenko, V.I.~Tretyak, Yu.G.~Zdesenko, Event generator DECAY4 for simulating double-beta processes and decays of radioactive nuclei, Phys. At. Nucl. 63 (2000)
 1282.
 \bibitem{DECAY0-prep} V.I.~Tretyak, in preparation.

\bibitem{Sonzogni:2002} A.A.~Sonzogni, Nuclear Data Sheets for A = 136, Nucl. Data Sheets 95 (2002) 837.

 \bibitem{Fel98} G.J.~Feldman, R.D.~Cousins, Unified approach to the classical statistical analysis of small signals, Phys. Rev. D 57 (1998) 3873.

  \bibitem{Krivoruchenko:2011} M.I.~Krivoruchenko et al., Resonance enhancement of neutrinoless double electron capture, Nucl. Phys. A 859 (2011) 140.

 \bibitem{Suhonen:2012} J.~Suhonen, Nuclear matrix elements for the resonant neutrinoless double electron capture, Eur. Phys. J. A 48 (2012) 51.

 \bibitem{Rumyantsev:1998} O.A.~Rumyantsev, M.H.~Urin, The strength of the analog and Gamow-Teller giant resonances and hindrance of the 2$\nu \beta \beta$-decay rate, Phys. Lett. B 443 (1998) 51.

 \bibitem{Abad:1984} J.~Abad, A.~Morales, R.~Nunez-Lagos, A.F.~Pacheco, An estimation of the rates of (two-neutrino) double beta decay and related processes, J. Physique 45 (1984) C3-147.

 \bibitem{Barea:2013} J.~Barea et al., Neutrinoless double-positron decay and positron-emitting electron capture in the interacting boson model, Phys. Rev. C 87 (2013) 057301.

 \bibitem{Suhonen:2003} J.~Suhonen, M. Aunola, Systematic study of neutrinoless double beta decay to excited 0$^+$ states, Nucl. Phys. A 723 (2003) 271.

\end{thebibliography}
\end{document}